\documentclass[aps,pra,reprint,reprintnumbers,amsmath,amssymb,superscriptaddress,longbibliography]{revtex4-1}
\usepackage{graphicx}
\usepackage{dcolumn}
\usepackage{colortbl}
\usepackage{enumitem}
\usepackage{placeins}
\def\up{\uparrow}
\def\down{\downarrow}
\def\ket#1{\left| #1\right\rangle}
\def\bra#1{\left\langle #1\right|}

\begin{document}

\title{Loschmidt echo driven by hyperfine and electric-quadrupole interactions\\ in nanoscale nuclear spin baths}

\author{Ekrem Taha  \surname{G\"{u}ldeste}}
\author{Ceyhun \surname{Bulutay}}
\email{bulutay@fen.bilkent.edu.tr}
\affiliation{Department of Physics, Bilkent University, Ankara 06800, Turkey}

\date{\today}

\begin{abstract}
The nuclear spin bath (NSB) dynamics and its quantum control are of importance for the storage 
and processing of quantum information within a semiconductor environment. In the presence of a carrier spin, 
primarily it is the hyperfine interaction that  rules the high frequency NSB characteristics. Here, we first 
study the overall coherence decay and rephasings in a hyperfine-driven NSB through the temporal and spectral behaviors 
of the so-called Loschmidt echo (LE). Its dependence on the NSB size, initial polarization, and coupling inhomogeneity 
are separately investigated, which leads to a simple phenomenological expression that can accommodate 
all of these attributes. Unlike the prevailing emphasis on spin 1/2, the NSBs with larger spin quantum numbers are 
equally considered. For this case, additionally the effect of nuclear electric quadrupole interaction is taken into 
account where its biaxiality term is influential on the decoherence. The insights gained from model systems 
are then put to use for two generic realistic semiconductor systems, namely, a donor center and a quantum dot that 
represent small and large nanoscale NSB examples, respectively. The spectrum of LE for large quantum dots can 
reach the 100~MHz range, whereas, for donor centers, it reduces to a few MHz, making them readily amenable for dynamical 
decoupling techniques. The effect of quadrupole interaction on LE is seen to be negligible for large quantum dots, 
while it becomes significant for donor centers, most notably in the form of depolarizing a polarized NSB. 
\end{abstract}


\maketitle

\section{Introduction}
Nuclear spins in a solid state matrix are largely immune to charge noise which grants them coherence lifetime 
in excess of one second at room temperature, which is by several orders longer than that of the electron spins \cite{maurer12}. 
This electrical isolation qualified them early on for several quantum information processing tasks \cite{kane98}. 
Generally, nuclear spins are thought to be ideal as quantum registers 
\cite{taylor03a,taylor03b,witzel07}, and in various ways they can be harnessed for gate operations 
\cite{taylor04,casanova17}. In virtually all of these cases, for rapid and convenient manipulation by electrical \cite{loss98} or optical 
\cite{imamoglu99} means an intermediary electron associated with a charged quantum dot or a defect center is exploited. 
For this composite system of an electron and the nuclear spin bath (NSB), the hyperfine (hf) interaction is the leading process that 
affects the coherence of both parties \cite{merkulov02,schliemann03,zhang06,latta11}. 

Intimately, this subject is linked with the central spin model that makes up a large body of literature \cite{gaudin76,barnes12,stanek13,stanek14}. 
Till now, the emphasis in this model has been on the decoherence of the electron, but undoubtedly it will be highly beneficial to 
visit the hf-driven coherence dynamics from the NSB standpoint.
One source of motivation for this comes from a proposal based on maximally entangled state generation between two electron spins 
by sequentially allowing them to interact with a NSB mediated by hf interaction (HFI) \cite{christ08}. Its noteworthy feature is 
that no information about the NSB, let alone a special preparation, is required for the success of this recipe.
As its offspring, a quantum interface between optical fields and the polarized nuclear spins was suggested for a singly 
charged quantum dot, again aided by HFI that allows both high-fidelity read-out and write-in of quantum information between the NSB 
and the output field \cite{schwager10}. 
The established  competence of dynamical decoupling techniques for suppressing the dipolar fluctuations is another point that 
needs to be reassessed in the case of the strong and inhomogeneous nature of the HFI \cite{suter16}.

These matters substantiate the need for a deeper understanding of the dynamics of NSB under HFI. For this purpose the Loschmidt 
echo (LE) is a suitable measure which corresponds to the return probability of spin bath to its initial state \cite{jalabert01}. 
It makes an ideal tool for tracking the bifurcated NSB dynamics when it is hf-coupled to an electron spin in a superposition state, 
i.e., a general qubit state \cite{yao07}. It needs to be mentioned that LE is directly accessible experimentally by means of 
nuclear magnetic resonance tools, where it has been used to monitor the degree of decoherence; see Refs.~\onlinecite{buljubasich15,sanchez16} 
and references therein. Quite recently it was employed for protecting fragile quantum superpositions in spin-1/2 clusters \cite{hahn17}, 
and for characterizing the spreading of initially localized quantum information across different degrees of freedom in many-body 
systems in the context of information scrambling \cite{chenu18}.

Our aim in this work is to develop a simple understanding of the HFI-driven NSB temporal and spectral characteristics, specifically 
by revealing the dependencies on bath size, coupling nonuniformity, initial state, and the nuclear spin quantum number $I$. 
We observe that, for unthermalized nanoscale NSBs having a narrow distribution, LE can reveal recoherence effects, as it resembles 
a closed system dynamics within the duration of interest \cite{yao07}. Remaining within the HFI-driven regime, under simple scalings 
we show that LE curves for different bath size or spin $I$ coalesce to a universal one. 

In the literature, exclusively the spin-1/2 NSB has been treated \cite{taylor03a,taylor03b,witzel07,taylor04,casanova17,imamoglu99,stanek13,stanek14,christ08,schwager10}, 
and sometimes indirectly through the so-called pseudospin approximation \cite{yao07}, despite the fact that group III-V semiconductors involve quadrupolar 
nuclei, where $I\ge 1$ \cite{levitt-book}. Therefore, we extend our consideration to the nuclear electric quadrupole interaction 
(QI) practically resulting from the atomistic strain in semiconductor structures for the case of quadrupolar 
NSB \cite{bulutay12,bulutay14}. We identify under what circumstances and how the action of QI on LE becomes significant. 
Finally, we consider two realistic cases of a lateral quantum dot that corresponds to a 
large NSB, and a donor center representing a small reservoir, and compare their LE spectra.

The paper is organized as follows. In Sec.~II we present our model Hamiltonian, a discussion of its connection to physical phenomena, 
and other theoretical elements that our analysis is based on. Section~III contains our results, where we first establish basic dependencies of 
the LE, and then use them in interpreting two realistic solid-state examples corresponding to small and large NSB prototypes. Conclusions 
are provided in Sec.~IV, and the Appendix embodies some derivations of LE expressions.

\section{Theory}
\subsection{Hyperfine interaction with the central spin}
The two sub-systems in our model are the central spin 1/2 (frequently referred to as the qubit) that is represented with 
the spin up/down basis states ($\ket{\up}$, $\ket{\down}$), and the homospin $I$ nuclei forming the bath sector. In this work
the considered nuclear spin length $I$ numerically ranges between 1/2 and 9/2. The qubit and bath spins together are treated 
in our model as a closed system within the timescale of relevance for the hf-dominant regime. In the absence of qubit longitudinal
relaxation, the so-called pure dephasing Hamiltonian \cite{giedke06} is expressed as
\begin{equation}
\label{H_hf}
\hat{H}=\hat{H}_+\otimes\ket{\up}\bra{\up} + \hat{H}_-\otimes\ket{\down}\bra{\down}\, ,
\end{equation}
where the nuclear spin dynamics is conditioned on the qubit states as 
$\ket{\up}\rightarrow\hat{H}_+$, $\ket{\down}\rightarrow\hat{H}_-$,  with
\begin{equation}
\label{H_pm}
\hat{H}_\pm=\pm\sum_i A_i \hat{I}^{z}_{i}\, .
\end{equation}
Here $\hat{I}^{z}_{i}$ is the $i$th nuclear spin operator's component along the central-spin quantization axis,
and $A_i$ is its hf coupling frequency. For convenience we set the Planck's constant to unity, $h \rightarrow 1$.

\subsection{Loschmidt echo}
To track its quantum coherence, we start the system with the qubit being in the superposition state in the chosen $z$-computational basis
$\ket{\psi} =C_+\ket{\up}+C_-\ket{\down}$ which is taken to be initially uncorrelated with the bath sector $\ket{B_0}$, 
hence in tensor product form
\begin{equation}
\ket{\Psi (t=0)} =\ket{\psi}\otimes\ket{B_0}\, .
\end{equation}
As the system evolves under the Hamiltonian of Eq.~(\ref{H_hf}) this product state turns into an entangled state,
\begin{equation}
\ket{\Psi (t)} =C_+\ket{\up}\otimes\ket{B_+(t)} +C_-\ket{\down}\otimes\ket{B_-(t)}\, .
\end{equation}
Therefore the initial superposition information of the qubit leaks to the bath state, which is a sign of loss of 
qubit coherence that can be identified from the degree of distinguishability of the two pathways from the bath sector as
\begin{equation}\label{Lbath}
L(t) =\langle B_-(t)|B_+(t)\rangle = \bra{B_0}e^{i\hat{H}_-t} e^{-i\hat{H}_+t}\ket{B_0}\, .
\end{equation}
This is directly related to the so-called the Loschmidt echo (LE), also known as the probability of return to initial configuration, as
$M(t)=\left | L(t)\right |^2$ \cite{jalabert01}.

For this essentially one-body Hamiltonian of Eq.~(\ref{H_pm}), an analytical form for $L(t)$ for a spin-$I$ environment can be written as,  
\begin{equation}\label{Lt}
L(t) =\prod_{i}\Big \lbrace\sum_{m_i=-I_i}^{I_i}W_i^{m_i}e^{-i2A_im_it}\Big \rbrace,
\end{equation}
where
\begin{equation}
W_i^{m_i}= \binom{2I_i}{I_i+m_i}   \left[\cos(\theta_i /2)\right]^{2(I_i+m_i)} \left[\sin(\theta_i /2)\right]^{2(I_i-m_i)}
\end{equation}
is the weight function, which is completely independent of azimuthal angle $\phi$, $m\in \{-I,-I+1,\dots,I-1,I\}$ 
are the possible eigenvalues along the quantization axis; $\theta$ is the polar angle; and the subscript $i$ again denotes 
the nuclear site index. The simplest case is the homospin-$1/2$ environment where Eq.~(\ref{Lt}) reduces to
\begin{equation}
L(t) =\prod_{i}\Big \lbrace \cos^2(\theta_i /2)e^{-iA_it} + \sin^2(\theta_i /2)e^{iA_it} \Big \rbrace,
\end{equation}
 which has a structure very similar to that derived in Eq.~(16) of \cite{cucchietti05}.
It is straightforward to calculate power spectra (see, the Appendix), $|M(f)|^2$ through 
the Fourier transform of LE, which yields
\begin{eqnarray}
M(f) & = & \sum\limits_{\substack{m_1,m_2,\dots,m_N,\\m_1',m_2',\dots,m_N'}}\Big( \prod_{i=1}^{N}W_i^{m_i}W_i^{m_i'} \Big) \nonumber\\
& & \times\delta\left(f +\dfrac{1}{\pi}\sum_i^{N}A_i( m_i-m_i')\right).
\end{eqnarray}

\subsection{Initial bath state}
For nanoscale spin baths, in contrast to mixed states the pure states become more appropriate and can be prepared 
through various means \cite{giedke06}. Moreover, the dependency on the initial nuclear spin states can be substantially 
suppressed by dynamical decoupling techniques \cite{suter16}. Therefore, we shall mainly employ different pure initial 
bath states, $\ket{B_0}$. For these, we assume the individual nuclear spins to be coherent 
spin states \cite{ma11} centered at the spherical angles $\ket{\theta_i, \phi_i}$ (see, the Appendix). For unpolarized baths we start 
with randomly selected angles from a uniform distribution over the Bloch sphere. In the case of baths with initial polarization, 
this distribution is restricted to a cone defined by a polar angle $\theta_p$.

\subsection{Nuclear electric quadrupole interaction}
As we mentioned in the Introduction we also consider nuclei with $I>1/2$, and they possess aspherical 
charge distributions giving rise to a nonzero electric quadrupole moment \cite{cohen57,das58}. These  
quadrupolar nuclei are affected by the gradient of an electric field that is present at a nuclear site. Such a setting 
becomes readily available in low-dimensional alloy structures of group III-V semiconductors (like InGaAs quantum dots) 
arising from the atomistic scale distortions within the tetrahedral bonding of polar constituents~\cite{bulutay12,bulutay14}.
Thus, a quadrupolar NSB has an additional interaction channel described by the Hamiltonian 
\begin{equation}
\label{H_q}
\hat{H}_Q = \sum_i\frac{f_{Qi}}{6}\left\{ 3\left(\hat{I}^z_i\right)^2  +  \frac{\eta_i}{2} 
\left[\left(\hat{I}^+_i\right)^2+\left(\hat{I}^-_i\right)^2\right] \right\}\, ,
\end{equation}
where $\hat{I}^{\pm}\equiv\hat{I}^x\pm i\hat{I}^y$ are the standard spin raising/lowering operators, $f_{Qi}$ and $\eta_i$ 
are respectively the quadrupolar frequency and the tensorial electric field gradient biaxiality at the $i$th nuclear site,  
and here we dropped a constant $\hat{I}_i^2$ term \cite{cohen57}. We should note that QI is not conditioned 
on the state of central spin, unlike the HFI. So, when both interactions coexist the total Hamiltonian
takes the form 
\begin{equation}
\label{H_full}
\hat{H}=\left(\hat{H}_Q+\hat{H}_+\right)\otimes\ket{\up}\bra{\up} + \left(\hat{H}_Q+\hat{H}_-\right)\otimes\ket{\down}\bra{\down}\, .
\end{equation}

\subsection{Physical relevance of the model}
Before proceeding further, we would like to express the particular timescale and physical context where this model 
is practically relevant. First of all, it should be mentioned that we are using the so-called Fermi-contact HFI
which applies to semiconductor conduction band electrons; for the holes the {\it dipolar} HFI needs to be considered 
\cite{stoneham-book,coish09}. In this work we are interested in the pure dephasing regime where no longitudinal relaxation 
hence qubit spin-flip takes place. This process becomes crucial in quantum information data writing stage where the 
spin flip-flop part of the HFI is set to resonance by applying a suitable external static magnetic field; but  during the 
storage it is intentionally detuned by turning off this field to maintain coherence 
\cite{deng05}. In general, for the so-called nuclear spin nonsecular part of the HFI, we refer to Cywi\'{n}ski et al. 
for conditions on how it can be ignored \cite{cywinski09}.

Furthermore due to fixed lattice spacing in a solid state environment that limits the proximity of two neighboring spins 
the dipole-dipole interaction typically proceeds within millisecond or longer durations \cite{casanova17}.
This makes it more than three orders of magnitude slower than that of the HFI and can be safely discarded within the
time frame under consideration here \cite{merkulov02,schliemann03}.
In connection to this, we should point out that we consider {\it a single realization} of NSB, as opposed to {\it ensemble averaged} 
calculations as in the seminal work of Ref.~\cite{merkulov02}. The latter particularly wipes out the coherent oscillations in LE by removing some 
of the integrals of motion which could be justified in the presence of dipole-dipole interaction \cite{erlingsson04}. As mentioned above, our work does 
not apply to this long-term regime.

Another related effect is the {\em indirect} HFI originating from the mean nuclear polarization which leads to additional 
nuclear spin precession \cite{merkulov10}. It plays a role in the relaxation of the electron spin transverse polarization 
\cite{yao06,deng06,witzel06}. However, since we focus on the NSB dynamics this indirect HFI can be practically omitted when 
the electron Knight field is high enough \cite{merkulov10}. 
Finally, the presence of an external magnetic field will result in a negligible Zeeman splitting in comparison to HFI 
because of the very small nuclear magnetic moment \cite{levitt-book}.
Under these conditions the Hamiltonian in Eq.~(\ref{H_full}) can serve as a good model for studying the evolution of NSB 
coherence subject to HFI of the central spin \cite{giedke06}, along with QI where applicable.

\section{Results}
\subsection{Study of basic dependencies}
We would like to gain a functional understanding of LE by probing separately the dependency on key variables before we confront 
realistic cases. In this section, we prefer to use normalized time and frequency, defined with respect to
the mean value of hf coupling constants, $\bar{A}=\sum_{i=1}^{N}A_i/N$, so that the normalized time becomes 
$\tilde{t}\equiv t \bar{A}$ and the normalized frequency is $\tilde{f}\equiv f/\bar{A}$.
We should note that other normalization schemes also exist in the literature \cite{stanek13,stanek14,hackmann14}.

\begin{figure}[ht]
		\begin{center}
		\includegraphics[width=1\columnwidth]{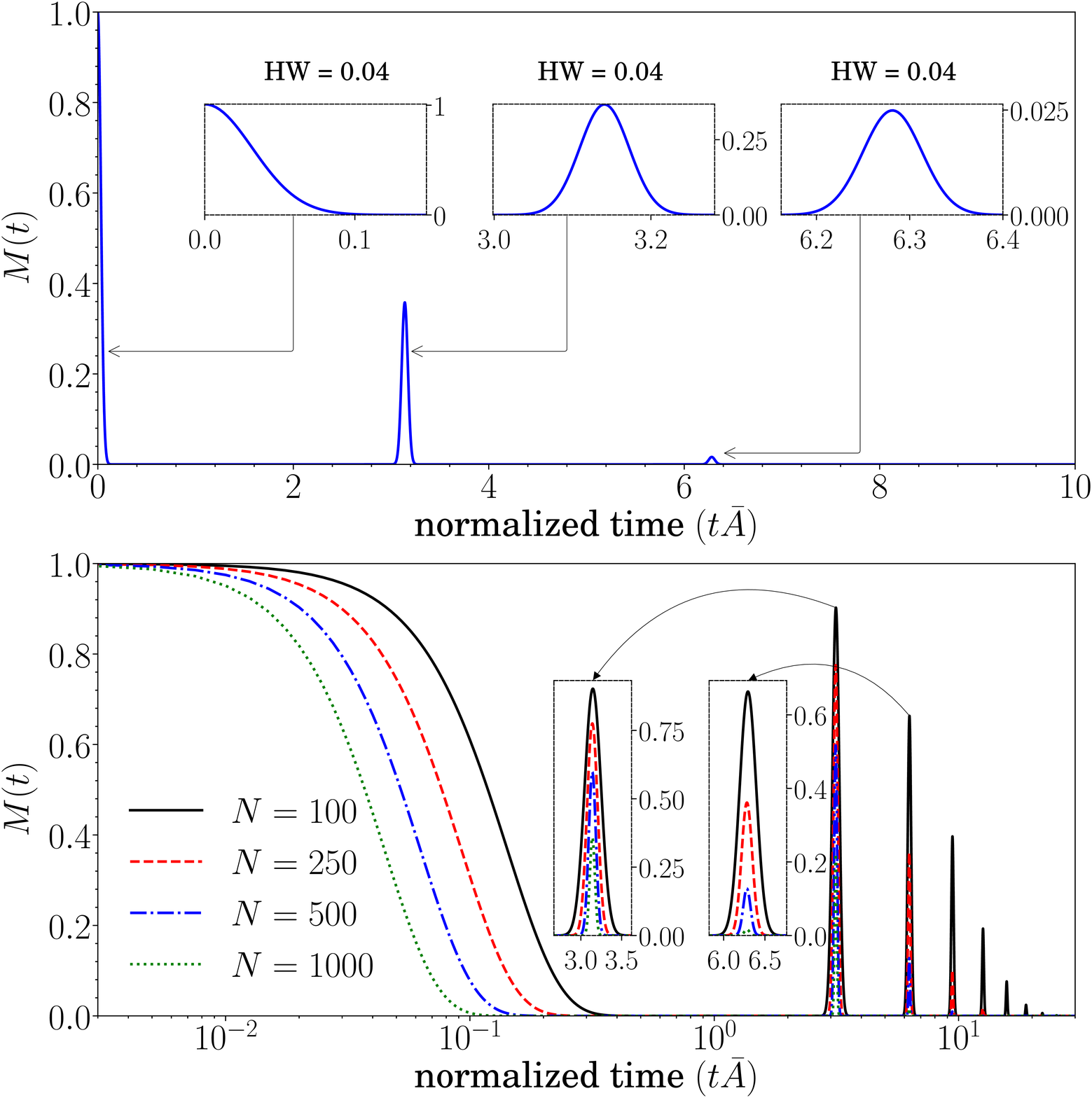}
		\caption{Top: LE for $N=1000$. Insets show the half-width (HW) of revivals. 
Bottom: Effect of different numbers of nuclear spins, $N$, forming the bath. In all cases $I=1/2$, 
$\Delta A_{\textit{max}} = 0.025\bar{A}$, and initial bath coherent spin states are uniformly distributed over the Bloch sphere.}
		\label{fig:A}
		\end{center}
		\end{figure}
	
We begin by analyzing a typical temporal behavior of LE of a spin-1/2 reservoir composed of $N=1000$ nuclei, each initially
starting as a coherent spin state $\ket{\theta_i, \phi_i}$ with the angles chosen randomly from a uniform distribution 
over the full Bloch sphere. The hf coupling constants of the spins in NSB are inevitably detuned from each 
other even for the homonuclear case due to spatial variation of the central electron wave function over the lattice. We assume a 
uniform spread in hf coupling constants with a maximum deviation of $0.025 \bar{A}$. As a matter of fact, this is quite 
small compared to actual cases, but our aim here is to demonstrate the level where it starts to inflict a significant effect.
Figure~\ref{fig:A} (top) shows the initial 
dephasing in LE followed by diminished-amplitude rephasings, all of which of the same Gaussian profile with equal halfwidths. 
The lower panel of Fig.~\ref{fig:A} displays the dependency of size $N$; as expected, larger NSB exhibits faster dephasing 
which is in agreement with the experimental observation that the decoherence rate increases with the number of dynamically 
coupled spins \cite{buljubasich15}. For a sufficiently large NSB (such as $N\geq1000$ here) these echos are periodic of the form 
$\left[\cos\tilde{t}\,\right]^{\alpha NI}$ with a Gaussian revival envelope.

\begin{figure}[ht]
		\begin{center}
		\includegraphics[width=1\columnwidth]{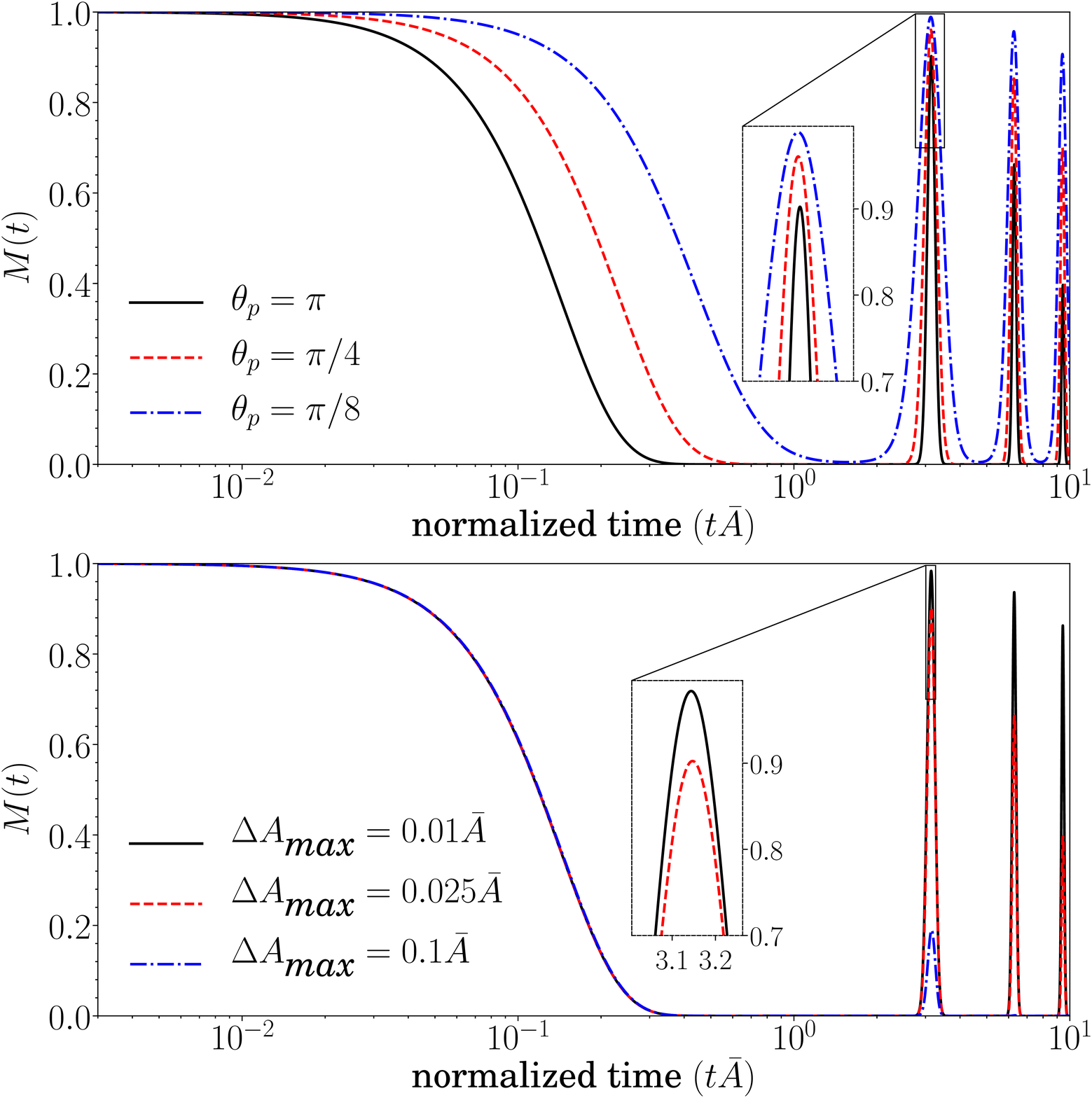}
		\caption{Top: Effect of initial nuclear spin polarization ($\theta_p$) on LE, $\Delta A_{\textit{max}} = 0.025\bar{A}$. 
Bottom: Effect of spread in the hf coupling constants  ($\Delta A_{\textit{max}}$) of 
individual nuclear spins; initial bath coherent spin states are uniformly distributed over the Bloch sphere. 
In all cases $N=100$, $I=1/2$.}
		\label{fig:B}
		\end{center}
		\end{figure}

Coherence time and the revival amplitudes highly depend on the initial bath polarization, which is illustrated on the upper plot 
of Fig.~\ref{fig:B}. Here, for each nuclear spin we choose an initial coherent spin state to be centered at the 
polar angle $\theta_i$ that is within a cone defined by the angle $\theta_p > \theta_i$ 
which approaches $\pi$ for the limit of unpolarized NSB (i.e., over the full Bloch sphere). Thus, this introduces a nonvanishing initial 
Overhauser field that persists in time within the pure dephasing model. As observed in this plot higher polarization 
of NSB results in increased echo amplitudes together with a wider halfwidth. In the lower part of Fig.~\ref{fig:B}, this time 
we study the effect of different spread of hf coupling constants, where 
$\Delta A_{\textit{max}}$ is the maximum deviation from the NSB mean value (\textit{i.e.,} $\Delta A_{\textit{max}} = \max\{|A_i-\bar{A}|\} $). 
The form $\left[\cos\tilde{t}\,\right]^{\alpha NI}$ deduced from Fig.~\ref{fig:A} implies that there should be no change in echo widths 
since the mean value of coupling constants ($\bar{A}$) remains same, which is indeed confirmed by this figure. 
Moreover, a narrower hf distribution causes larger rephasing amplitudes.

\begin{figure}[ht]
		\begin{center}
		\includegraphics[width=1\columnwidth]{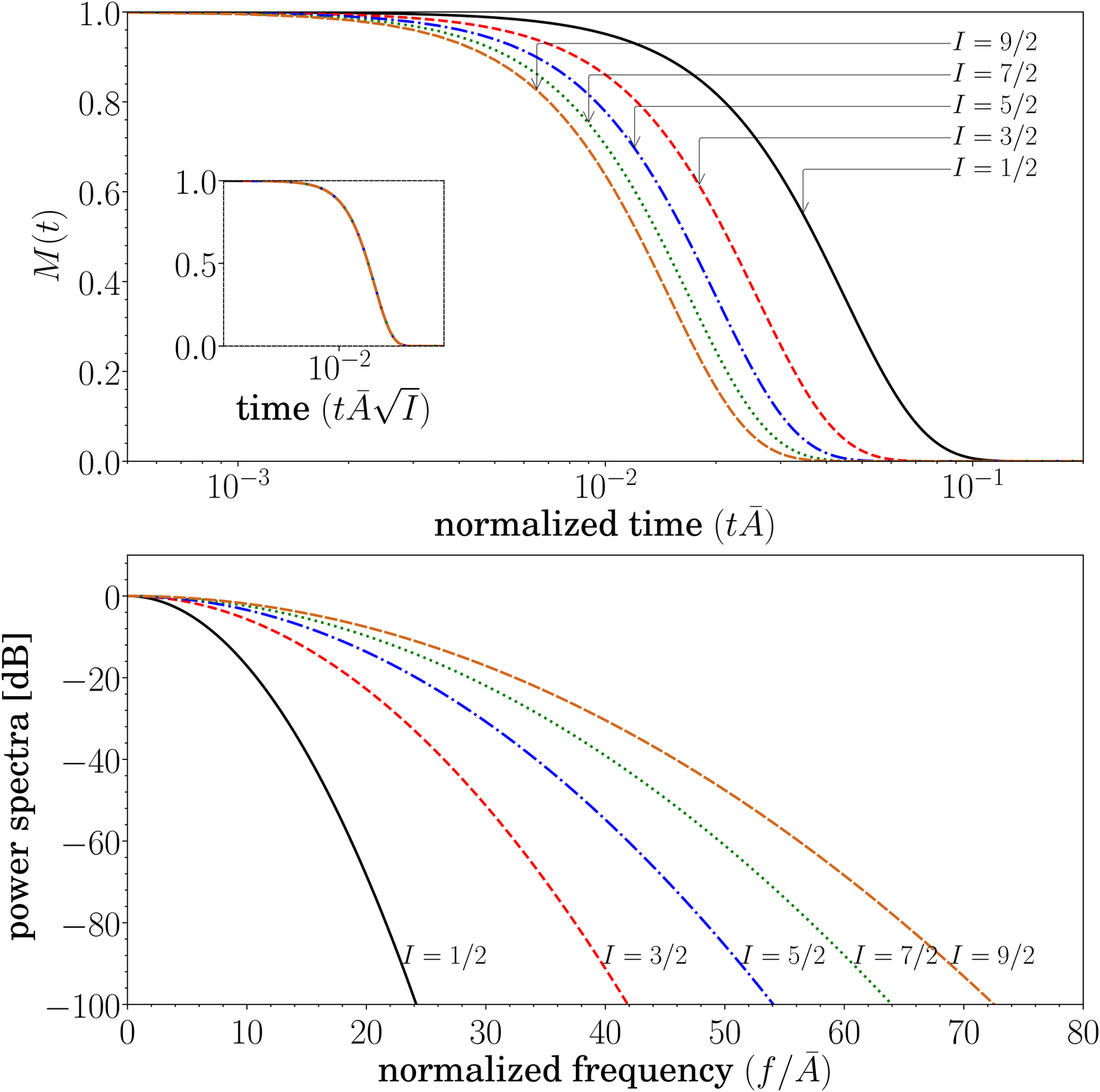}
		\caption{Comparison of different spin-$I$ values. Top: Temporal behavior; inset illustrates the coalescence of the 
		family of curves under the indicated normalization. Bottom: Spectral behavior. In all cases $N=1000$, 
		$\Delta A_{\textit{max}} = \bar{A}$, and the initial bath spins are uniformly distributed over the Bloch sphere.}
		\label{fig:C}
		\end{center}
		\end{figure}

Next, we study the dependency on the nuclear spin quantum number $I$. As this value is increased from 1/2 up to 9/2 its 
eigenspectrum gets denser, and more closely resembles a classical spin when $I\gg 1$ \cite{petrov12,opatrny15}. 
Figure~\ref{fig:C} (top) shows that the LE width decreases with increasing $I$, as it directly increases the mean hf 
field, causing faster nuclear spin precession, and hence faster dephasing. These family of curves coalesce to a single one 
(see the inset of Fig.~\ref{fig:C}) under the $t\bar{A}\sqrt{I}$ time scaling. The lower panel in Fig.~\ref{fig:C} displays the corresponding 
power spectra of the temporal behavior. As our primary aim is to compare spectral broadenings under various cases, here and throughout this
work each spectrum is vertically shifted to set its dc limit to $0$~dB. This facilitates verifying the widening of LE spectrum in proportion to 
$\sqrt{I}$ for homospin-$I$ NSBs.

An analytical derivation of LE for a general NSB would be highly desirable, but it has remained a formidable task. 
Cucchietti {\textit et al.} obtained a form valid under restrictive assumptions and only for spin-1/2 baths \cite{cucchietti07}. 
On the other hand our controlled-parameter studies as summarized in Figs.~\ref{fig:A}-\ref{fig:C} lead to a widely applicable 
phenomenological expression given by
\begin{equation}
\label{fit}
M(\tilde{t}) \sim \exp\left[ -NI \left( \alpha_p \sin^2(\tilde{t})+\beta_p\sigma^2\tilde{t}\,^2\right)\, \right] ,
\end{equation}
where $\sigma^2$ is the variance of the hf coupling constants and $\alpha_p$, $\beta_p$ are NSB polarization-dependent 
fitting parameters. It faithfully captures all of the size, spin-$I$, and hf coupling inhomogeneity dependencies of both echo
periodicity and amplitudes for $N\gtrsim 1000$ NSBs, and has a Gaussian form. Specifically, it predicts 
that the inhomogeneous broadening in hf couplings has no effect on the initial coherence decay rate 
(see the bottom panel in Fig.~\ref{fig:B}), which is instead controlled by the $NI$ product, together with the initial 
bath polarization as shown in Figs.~\ref{fig:A}-\ref{fig:C}.
	
\begin{figure}[ht]
		\begin{center}
		\includegraphics[width=1\columnwidth]{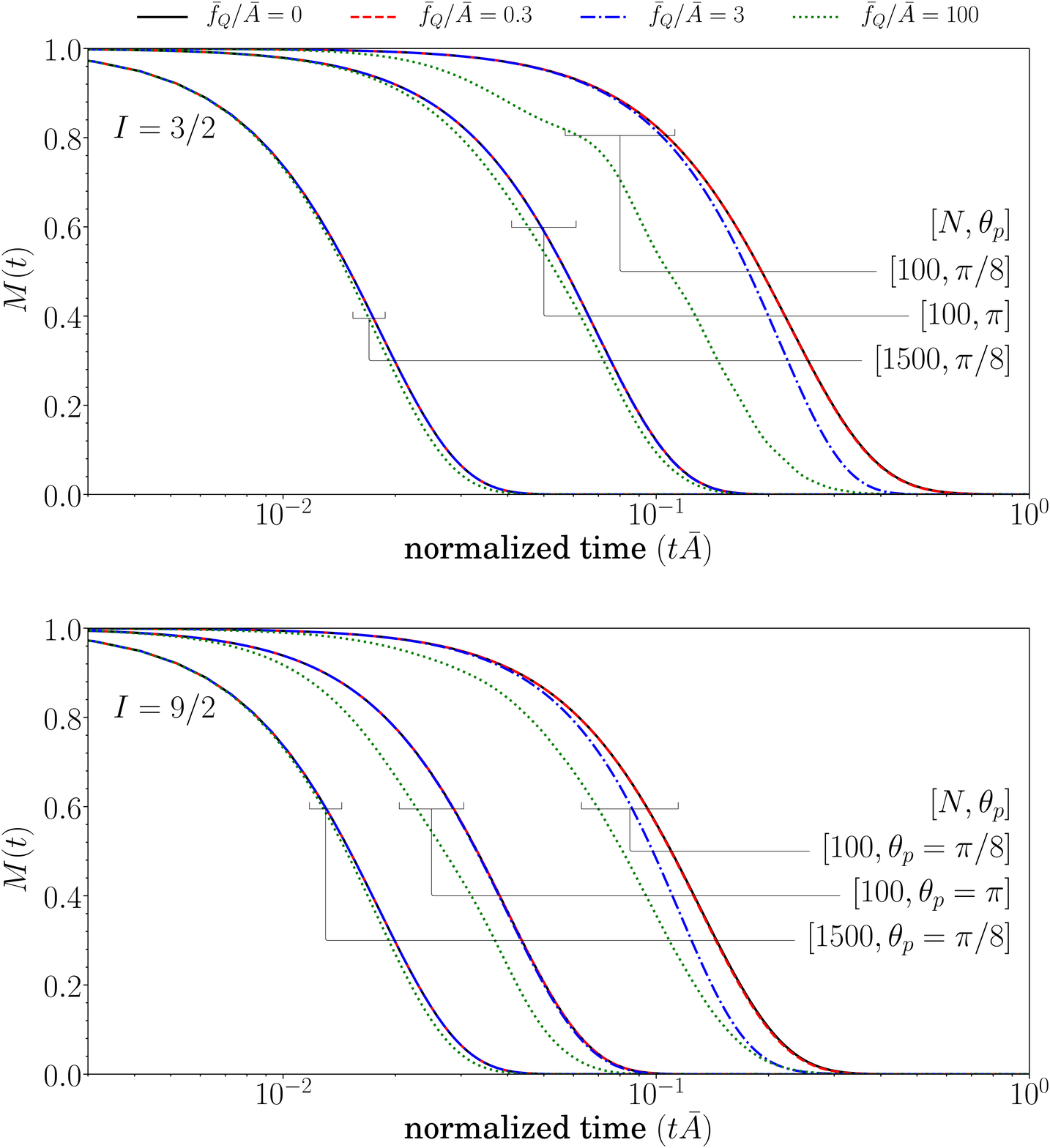}
		\caption{Effect of QI on unpolarized $(\theta_p=\pi)$ and polarized $(\theta_p=\pi/8)$ NSBs with 
                (top) $I=3/2$ and (bottom) $I=9/2$, for two different $N$ values with $\Delta A_{\textit{max}} = \bar{A}$.}
		\label{fig:D}
		\end{center}
		\end{figure}

So far, we have only included the hf coupling of each nucleus with the central spin [Eq.~(\ref{H_hf})]. 
In the case of quadrupolar NSBs having $I\ge 1$ the QI as described by Eq.~(\ref{H_q}) becomes operational. 
In Fig.~\ref{fig:D} the temporal behaviors of LE of spin-$3/2$ and -$9/2$ NSBs are compared for various 
mean $\bar{f}_Q=\sum_{i=1}^{N}f_{Qi}/N$ rates from weak to strong coupling limits. We should point out that 
the QI has a null effect on LE for a nuclear spin under $\eta_i=0$, i.e., at a {\em uniaxial} electric field 
gradient site \cite{bulutay14}. This is because the $(\hat{I}_i^z)^2$ term 
in Eq.~(\ref{H_q}) commutes with the $\pm \hat{I}_i^z$ parts of HFI; that is, the fluctuations caused by $(\hat{I}_i^\pm)^2$
terms are critical, and together with them, the $(\hat{I}_i^z)^2$ term imposes a nontrivial outcome on the 
dynamics. This necessitates $\eta>0$, where for alloy quantum dots (like In$_x$Ga$_{1-x}$As), 
$\eta\sim 0.2-0.6$ \cite{bulutay12}. Since $\eta_i$ term appears in the product with $f_{Qi}$ in Eq.~(\ref{H_q}), 
for simplicity we fix the former to $\eta_i=0.5$ for all nuclear spins, and let $\Delta f_{Q,\textit{max}}=0.2\bar{f}_Q$.    
The distribution of hf coupling constants is taken as $\Delta A_{\textit{max}} = \bar{A}$, which prohibits any revival 
of LE beyond the initial decay as inferred from Fig.~\ref{fig:B}. We can mention that its precise value is not critical as a 
choice of, say, $\Delta A_{\textit{max}} = 0.25 \bar{A}$ generates indiscernible results within our time frame of interest. 
In such a practical setting, we first observe that  
for a given bath size $N$, as QI gets stronger it causes a faster decay, and hence broadens the frequency spectrum of LE. 
Moreover, the contribution of QI is much more pronounced on polarized NSBs (minding the logarithmic timescale in 
Fig.~\ref{fig:D}), acting in the direction to depolarize NSB. Furthermore, we note that the significance of QI decreases as 
the bath size $N$ increases. This stems from the fact that the (normalized) first decay rate $\tilde{f}_{1D}$, as can be 
extracted from the variance of $M(\tilde{t})$ from Eq.~(\ref{fit}), has the dependence $\tilde{f}_{1D}\propto\sqrt{NI}$, so 
that, for a given $\bar{f}_Q$, as $N$ increases so does $\tilde{f}_{1D}$, rendering ineffective the QI within the 
first decay time frame of the LE.

\subsection{Realistic solid-state models}
In the light of these basic findings we are ready to compute and interpret LE of realistic NSBs, for which we choose a 
donor/defect center within a semiconductor host matrix, and a lateral quantum dot, to represent 
small and large reservoir cases, respectively. For the spatial distribution of hf coupling constants the electron envelope 
wave function is chosen to be of the form \cite{hackmann14}
\begin{equation}
\Psi(r_i)=\Psi(0) \exp \left(  -\dfrac{r_i^2}{2l_0^2}\right),
\end{equation}
where $r_i$ is the distance of the $i$th nuclear site from the origin and $l_0$ is the electron confinement radius.
In our choice, the NSB constitutes all the nuclei with $|\Psi(r_i)/\Psi(0)|>10^{-3}$.
An effective number of spins $N_{\textit{eff}}$ can be defined as \cite{schliemann03}
\begin{equation}
N_{\textit{eff}}=\rho\dfrac{4\pi l_0^3}{3v_0}\, ,
\end{equation}
in terms of the ratio of spinful nuclei, $\rho$,  and  the volume occupied by a single atom, $v_0$, constrained by normalization 
condition $v_0\sum_i|\Psi(r_i)|^2\approx 1$. 
For a typical donor center with a radius of 5~nm, the number of effective spins 
is $N_{\textit{eff}}=100$, and the sum of coupling constants becomes  $\sum_{i=1}^{N_{\textit{eff}}} A_i \approx 0.141~\mu$eV 
for the ratio of $\rho\approx 0.05$ of spinful nuclei, as in silicon \cite{coish04}. In the case of a large NSB, we choose a disk-shaped 
quantum dot where the electron envelope wave function is taken to be Gaussian (uniform) in the radial (growth) direction, with radius (height)
12.5~nm (3~nm). The effective number of spins becomes $N_{\textit{eff}}=10~000$,
and the sum of couplings is estimated as $\sum_{i=1}^{N_{\textit{eff}}} A_i \approx 70.856~\mu $eV.

\begin{figure}[ht]
		\begin{center}
		\includegraphics[width=1\columnwidth]{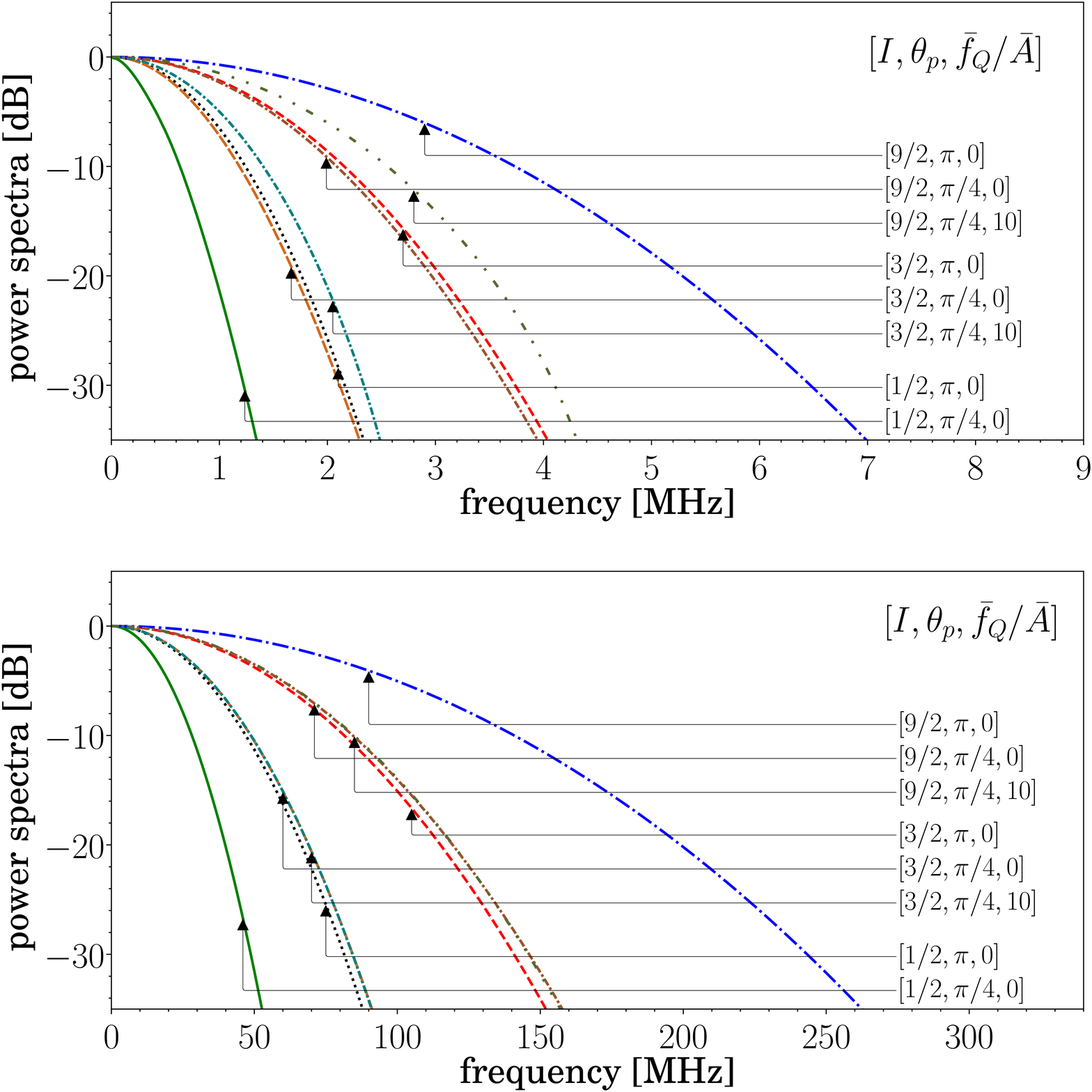}
		\caption{Power spectra of LE for realistic systems under different spin-$I$, polarization ($\theta_p$) 
		and quadrupolar frequencies ($\bar{f}_Q$). Top: donor center, $N_{\textit{eff}}=100$. Bottom: 
		Lateral quantum dot, $N_{\textit{eff}}=10~000$. For the bottom case, $\bar{f}_Q/\bar{A}$=0, 10 curves 
		become indiscernible for each $I$.}
		\label{fig:E}
		\end{center}
		\end{figure}

The LE power spectra for both systems are compared in Fig.~\ref{fig:E} under various parameters which corroborates the individual
traits discussed in the previous section. First of all, a finite initial polarization of the NSB significantly narrows 
the spectrum compared to unpolarized one. Moreover, as observed in Fig.~\ref{fig:C}, there occurs $\sqrt{I}$ widening of the 
spectra for spin-$I$ NSBs.  Apart from these common features, the generic quantum dot system has about two orders of magnitude 
broader frequency bandwidth compared to the donor center case with the latter being limited to a few megahertz. This directly follows 
from their $\sqrt{N_{\textit{eff}}}$ ratio, as demonstrated in the lower part of Fig.~\ref{fig:A} and Eq.~(\ref{fit}). Hence, for 
a spin-9/2 quantum dot (as with indium nuclei) the power spectrum can spread to some 100~MHz. Regarding QI, the quadrupolar 
frequency dictated by strain is typically in the range $f_Q\sim$ 2--8~MHz for typical quantum dots \cite{bulutay12}, and 3--6~MHz 
for defect centers, as in hexagonal BN flakes \cite{lovchinsky17}. In our examples here, the mean hf coupling constant, $\bar{A}$ 
is about 0.34~MHz (1.7~MHz) for the donor center (quantum dot), so as a representative value we consider $\bar{f}_Q/\bar{A}=10$,
along with $\eta_i=0.5$. From Fig.~\ref{fig:E} it can be seen that QI is ineffective on LE for a large quantum dot, whereas it 
has an influence on the donor center with polarized NSB having a small $NI$ product, in line with our conclusions from 
Fig.~\ref{fig:D} and Eq.~(\ref{fit}).

Finally, we would like to comment on the utility of such power spectra as in Fig.~\ref{fig:E}. In simple terms,
they specify the characteristic bandwidth of HFI and QI fluctuations in relation to the qubit coherence. 
As such, this may help to assess the efficacy of the dynamical decoupling techniques \cite{suter16}.
In a more specific context, the spectrum of NSB hf fluctuations plays a crucial role in the recently discovered
hf-mediated electric dipole spin resonance, in the form of both driving and detuning it \cite{laird07,rashba08}.

\FloatBarrier
\section{Conclusions}
HFI is commonly the dominant process that governs the short-term dynamics of the NSB in solid
state systems \cite{schliemann03}. The analysis of the hf-induced quantum fluctuations can be worthwhile for
various practical settings, such as prolonging the qubit coherence in the storage 
phase of quantum registers \cite{deng06,yang17}, or for obtaining indistinguishable photons 
from quantum dots having a resident electron \cite{malein16}. This work offers a simple theoretical exposition 
via the temporal and spectral characterization of the LE which is an experimentally measurable correlation for the 
degradation of the information contained in a quantum state in nanoscale NSBs \cite{buljubasich15,sanchez16}.
We extract basic dependencies on various reservoir parameters like size, initial polarization, coupling inhomogeneity, and
spin quantum number, and we suggest a phenomenological LE expression. We hope that it may also initiate further 
theoretical studies for its rigorous derivation.

Additionally, the effect of QI on LE is taken into account for the quadrupolar nuclei which are prevalent in III-V semiconductors.
In particular, it is the QI biaxiality term that has important ramifications on the qubit decoherence.
From the moderate coupling regime onwards ($\bar{f}_Q \gtrsim \bar{A}$) QI causes a faster decay of initial coherence that gets
more pronounced for polarized and small $NI$-product NSBs. Lastly, we contrasted two realistic cases of a donor center and a quantum 
dot representing small and large NSBs, respectively. Here, for quantum dots with $NI \gtrsim 5000$, the LE spectrum can stretch 
to 100 MHz range, and the effect of QI is rather negligible. On the other hand for donor centers, as this width narrows down by 
more than an order of magnitude the dynamical decoupling techniques become feasible, and at the same time QI can show its influence.

Throughout our work we excluded the intrabath interactions which come into play in the lower frequency regime. To extend it to 
this long-term dynamics where new experimental findings are available \cite{stanley14}, or alternatively  to study spin diffusion 
phenomena \cite{gong11}, efficient many-body techniques specifically devised for handling a large number of spins, like the
cluster-correlation expansion, can be invoked \cite{yang08,yang17}.

\begin{acknowledgments}
We thank Mustafa Kahraman for useful discussions on numerics. 
This work was supported by T\"UB\.ITAK, T\"{u}rkiye Bilimsel ve Teknolojik Ara\c{s}tirma Kurumu through Project No. 114F409. 
\end{acknowledgments}

\appendix

\section*{Appendix: Some expressions on LE}
A coherent spin state $\ket{\Omega =(\theta,\phi) }$ can be expressed as,
\begin{eqnarray}
\ket{\Omega} & = & \sum_{m=-I}^{m=I}\binom{2I}{I+m}^{1/2}\left[\cos(\theta/2)\right]^{I+m} \nonumber \\ 
& & \times \left[\sin(\theta/2)\right]^{I-m}e^{-i(I-m)\phi}\ket{m}.
\end{eqnarray}
Initially at $t=0$, the overall bath state can be expressed as the tensor product of spin coherent states, meaning for this one-body 
Hamiltonian we can compute $L(t)$ as the product of individual spin evolutions. Then, Eq.~(\ref{Lbath}) becomes
\begin{equation}
L(t)=\prod_{i=1}^{N}\bra{\Omega_i(0)}e^{-i2A_i\hat{I}_i^zt}\ket{\Omega_i(0)},
\end{equation}
from which we can directly arrive at
\begin{equation}\label{individual}
L(t) =\prod_{i=1}^{N}\Big \lbrace\sum_{m_i=-I_i}^{I_i}W_i^{m_i}e^{-i2A_im_it}\Big \rbrace
\end{equation}
after carrying out inner products.
We can rewrite Eq.~($\ref{individual}$) as
\begin{equation}
L(t) = \prod_{i=1}^{N} L_i(t),
\end{equation}
where,
\begin{equation}
L_i(t)=\sum_{m_i=-I_i}^{I_i}W_i^{m_i}e^{-i2A_im_it}.
\end{equation}

The Fourier transform of $L(t)$ becomes the convolution of all $L_i(f)$ in the frequency domain in the form
	\begin{equation}\label{convol}
	L(f) = L_1(f) * L_2(f) * \cdots *L_N(f).
	\end{equation}	
Then, calculating $L_i(f)$ yields
	\begin{align}
	L_i( f) &= \int_{-\infty}^{\infty} e^{-i2\pi  f} \sum_{m_i=-I_i}^{I_i}W_i^{m_i}e^{-i2A_im_it} dt \, , \notag\\
	&=\sum_{m_i=-I_i}^{I_i}W_i^{m_i}\int_{-\infty}^{\infty}e^{-i2 \pi  f}e^{-i2A_im_it} dt \, ,\notag \\
	&= \sum_{m_i=-I_i}^{I_i}W_i^{m_i} \delta(f+A_i m_i/\pi).
	\end{align}
Inserting  this expression into Eq.~(\ref{convol}) leads to
 \begin{equation}
 L(f) = \sum_{m_1,m_2,\dots,m_N} \Big( \prod_{i=1}^{N}W_i^{m_i} \Big)\delta \left(f +\dfrac{1}{\pi}\sum_i^{N}m_iA_i\right).
 \end{equation}
 Similarly for its complex conjugate we have
 \begin{equation}
 \left[L(f)\right]^* = \sum_{m'_1,m'_2,\dots,m'_N} \Big( \prod_{j=1}^{N}W_j^{m'_j} \Big)\delta\left( f -\dfrac{1}{\pi}\sum_j^{N}m_j A_j\right).
 \end{equation}
Hence, the Fourier transform of LE, $M(f) =  L(f)* \left[L(f)\right]^*$, is given by the expression
\begin{eqnarray}
M(f) & = & \sum\limits_{\substack{m_1,m_2,\dots,m_N,\\m_1',m_2',\dots,m_N'}}\Big( \prod_{i=1}^{N}W_i^{m_i}W_i^{m_i'} \Big) \nonumber\\
& & \times\delta\left[f +\dfrac{1}{\pi}\sum_i^{N}( m_i-m_i')A_i\right].
\end{eqnarray}

\end{document}